\documentclass[twocolumn,final]{IEEEtran}
\usepackage{amsthm,amssymb,graphicx,multirow,amsmath,color,amsfonts}
\usepackage[update,prepend]{epstopdf}
\usepackage[noadjust]{cite}
\usepackage{paralist}
\usepackage[latin1]{inputenc}

\pdfoutput=1

\def\E{\mathbb{E}}

\def\P{\mathbb{P}}
\def\pc{\mathtt{P_c}}

\def\R{\mathbb{R}}

\def\dd{\mathrm{d}}
\def\T{\beta}							% Threshold = \beta_i
\def\sir{\mathtt{SIR}}

\def\calE{\mathcal{E}}

\newtheorem{lemma}{Lemma}{}
  \newtheorem{thm}{Theorem}

  \newtheorem{cor}[thm]{Corollary}\vspace{-1in}

	\newtheorem{remark}{Remark}

\allowdisplaybreaks % Allows breaking of eqnarray over multiple pages (avoids unnecessary blanks in the document before eqnarray)
\begin{document}
\graphicspath{{./Figures/}}
\title{Modeling Non-Uniform UE Distributions in Downlink Cellular Networks
}
\author{Harpreet S. Dhillon, Radha Krishna Ganti and Jeffrey G. Andrews
\thanks{H. S. Dhillon and J. G. Andrews are with the Wireless Networking and Communications Group (WNCG), The University of Texas at Austin (email: dhillon@utexas.edu and jandrews@ece.utexas.edu). R. K. Ganti is with the department of EE at Indian Institute of Technology Madras (email: rganti@ee.iitm.ac.in). \hfill Manuscript updated: \today.} 
}

\maketitle
\thispagestyle{empty}
\pagestyle{empty}

\begin{abstract}
A recent way to model and analyze downlink cellular networks is by using random spatial models. Assuming user equipment (UE) distribution to be uniform, the analysis is performed at a typical UE located at the origin. While this method of sampling UEs provides statistics averaged over the UE locations, it is not possible to sample cell interior and cell edge UEs separately. This complicates the problem of analyzing deployment scenarios involving non-uniform distribution of UEs, especially when the locations of the UEs and the base stations (BSs) are dependent. To facilitate this separation, we propose a new tractable method of sampling UEs by {\em conditionally thinning} the BS point process and show that the resulting framework can be used as a tractable generative model to study cellular networks with non-uniform UE distribution.
\end{abstract}

\section{Introduction}
A promising new way to model cellular networks is by using random spatial models, where the BS locations are assumed to form a realization of some spatial point process, typically the Poisson Point Process (PPP). Modeling the UE locations as an independent PPP, the downlink analysis is performed at a typical UE assumed to be located at the origin. Owing to its tractability, this  model leads to simple closed form expressions for key metrics such as coverage and average rate over the entire network~\cite{AndBacJ2011,DhiGanJ2012,MukJ2012}. Although it provides a way to study average statistics over the complete network, it does not provide any handle on studying the performance of cell edge and cell interior UEs separately. Consequently, it is not possible to model deployment scenarios involving non-uniform UE distribution, especially when the UE and BS locations are dependent.  Naturally, such flexibility is desirable, especially to study the current capacity-centric deployments where BSs are specifically deployed to be proximate to areas of high UE density. The most popular and perhaps the only available option to handle such scenarios is through detailed system level simulations, which are both time consuming and have to be focused on a limited range of system parameters~\cite{GanJosC1997,StoVisC2009}.

As a first step towards a tractable model, we propose a slight modification in the way UEs are sampled by introducing the idea of {\em conditional thinning}. Starting with a higher density of BSs, we first assume that a typical UE is located at the origin. After selecting the serving BS, we condition on this active link and independently thin the rest of the BS point process so that the resulting density matches the desired density of the actual BSs. The thinning operation pushes the typical UE in the cell interior relative to the new cell edge defined by the resulting point process. The bias induced in the location of a typical UE towards its serving BS can be tuned by varying the thinning probability. We make this notion precise by deriving the distribution of the ratio of the distances of the UE to its serving BS and the dominant interferer as a function of the thinning probability. We also show that this framework can be used as a tractable generative model to study non-uniform UE distributions where the UEs are more likely to lie closer to the BS. The exact analysis of such non-uniform UE distributions is in general hard due to the correlation present in the UE and BS locations. The impact of the proposed model on the cellular performance analysis and its key differences from the existing model based on uniform UE distribution are highlighted in terms of the coverage predictions.

It is worth noting that although this work is developed in the context of cellular networks, it applies to much wider class of point process problems involving dependence in the location of the observation point and the point process.

\section{Proposed Method of Sampling UEs}
We consider a homogeneous PPP $\Phi$ of density $\lambda$, a thinned version of which will eventually model the BS locations. %The parameter $p \in (0, 1]$ is the thinning probability and controls the extend to which a typical UE is pushed to the cell interior, as explained later in this section.
The downlink analysis is performed at a typical UE assumed to be at the origin~\cite{StoKenB1995}. The received power at a typical UE from the BS located at $x \in \Phi$ is 
\begin{align}
P_{x} = P h_{x} \|x\|^{-\alpha},
\end{align}
where $P$ is the transmit power, $h_{x} \sim \exp(1)$ models channel power distribution under Rayleigh fading and $\|x\|^{-\alpha}$ models standard distance based path loss with $\alpha>2$ being the path loss exponent. More general fading distributions can be studied using tools developed in~\cite{BacMuhJ2009,MadResC2009}. For simplicity of exposition, we will ignore thermal noise in this discussion.

\begin{figure*}[t]
\centering
\includegraphics[width=0.32\textwidth]{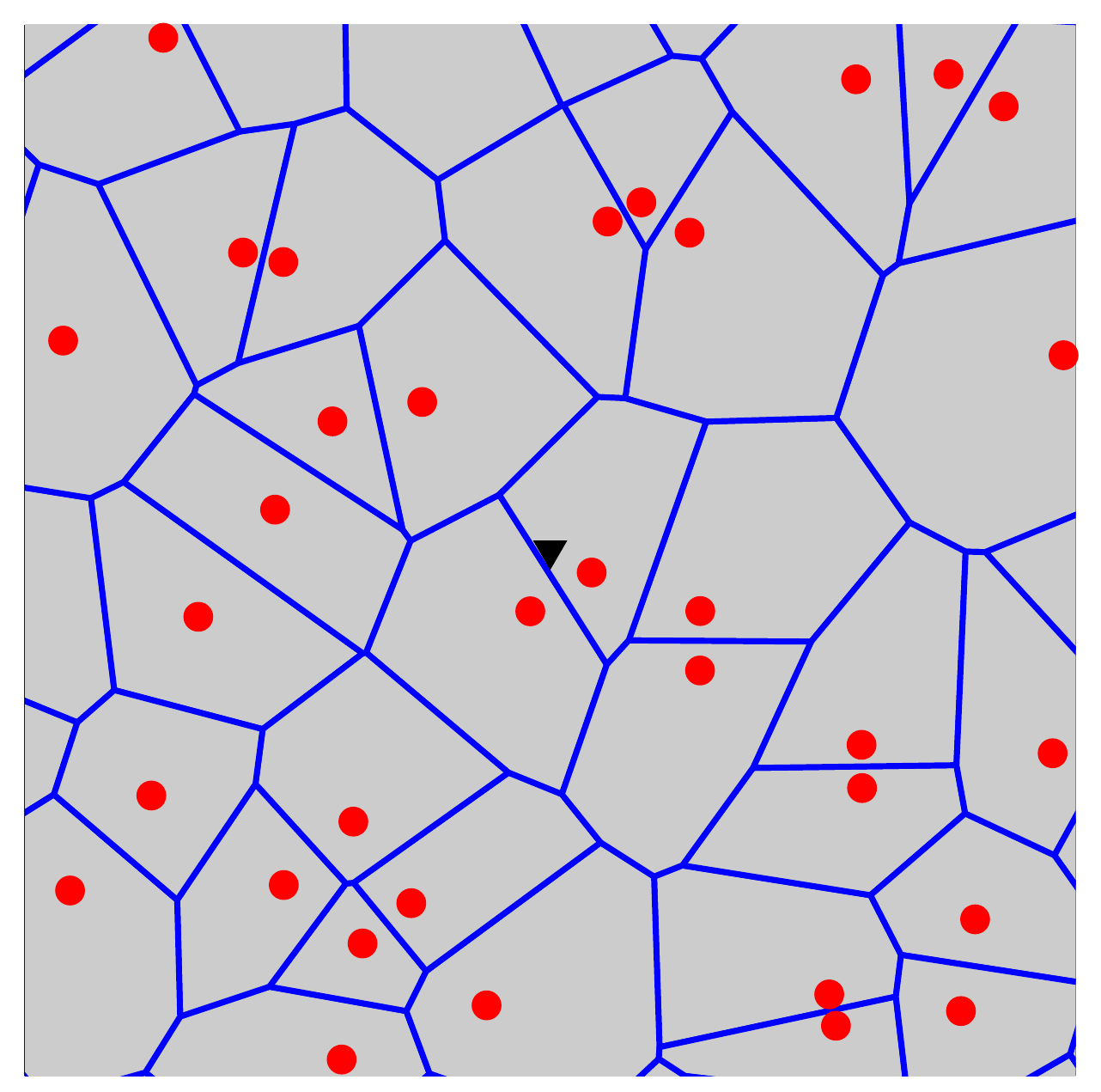}
\includegraphics[width=0.32\textwidth]{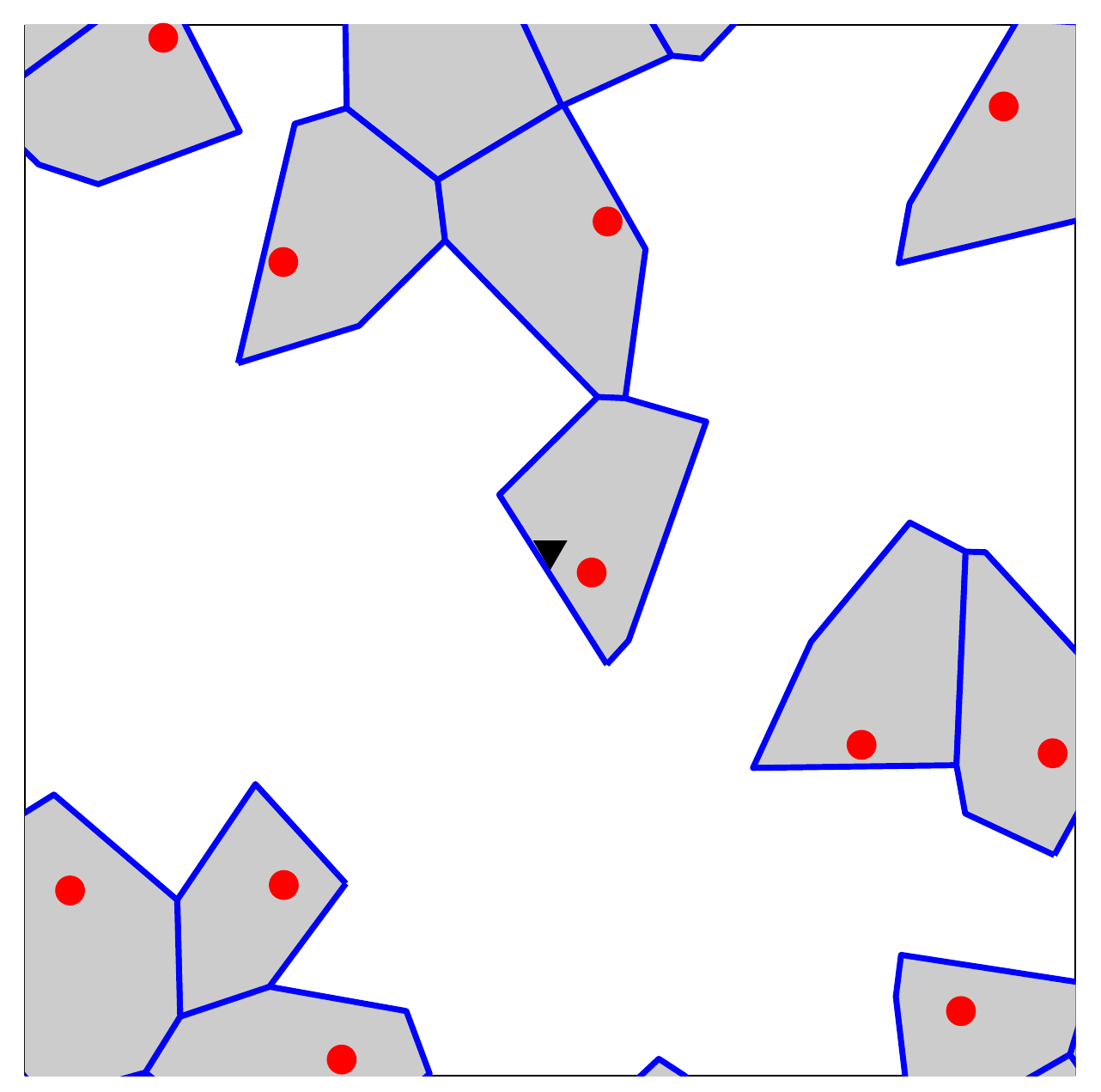}
\includegraphics[width=0.32\textwidth]{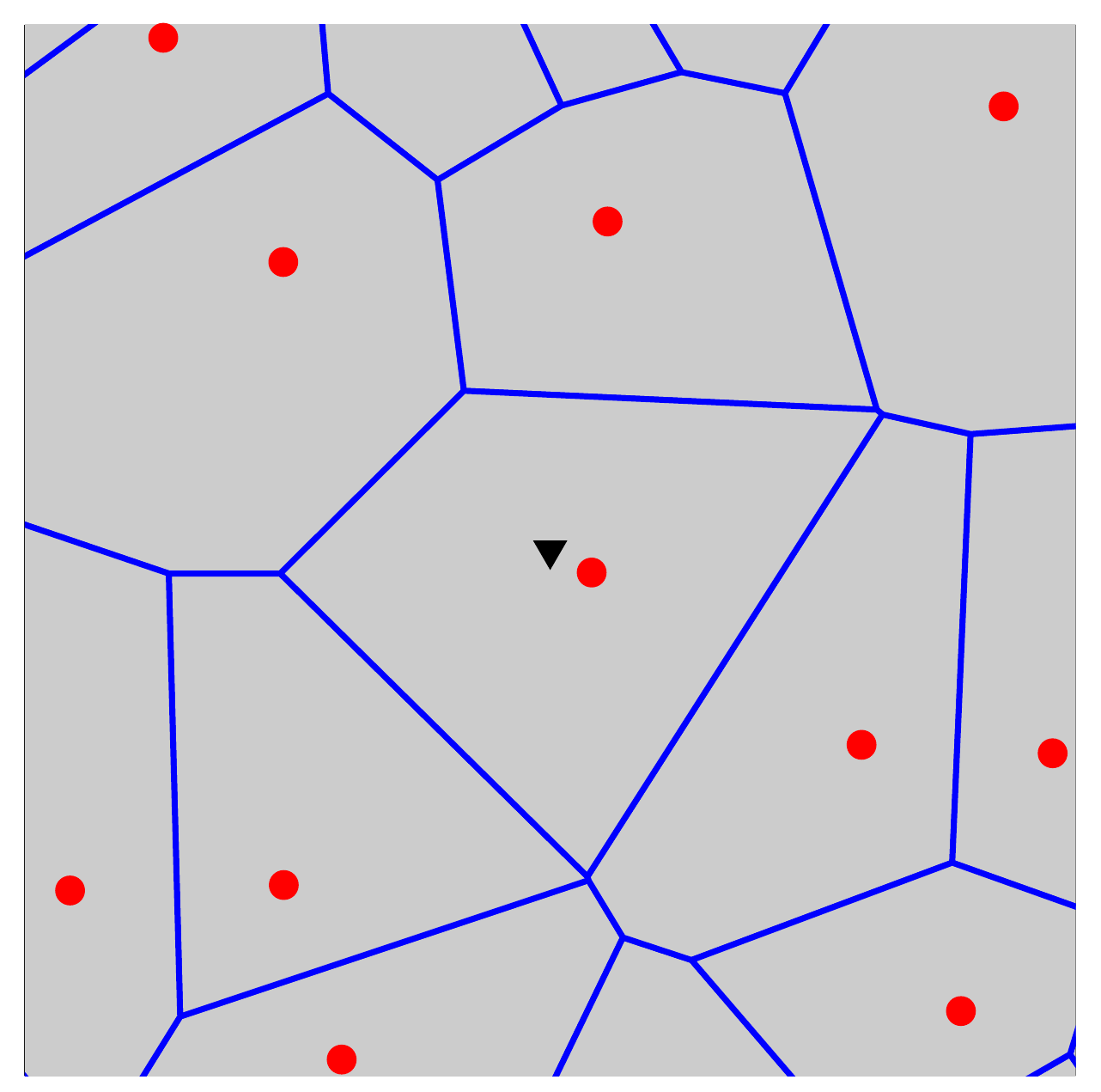}
\caption{(left) The Voronoi tessellation of the point process $\Phi$. Dark triangle denotes a typical UE. (middle) The point process thinned by $p = .3$. The remaining points form BS point process $\Phi_b$. (right) The Voronoi tessellation of $\Phi_b$. Observe that the typical point is now in the cell interior.}
\label{fig:1tier1}
\end{figure*}

The proposed method of sampling the interior UEs can be understood in two simple steps shown in Fig.~\ref{fig:1tier1}. The first step is to identify the serving BS in $\Phi$ for a typical UE depending upon the cell association technique being considered. While the proposed method is general, for brevity we consider nearest-neighbor cell association model discussed in~\cite{AndBacJ2011}, where each UE connects to its nearest BS. It is also the same as maximum average power connectivity model, where each UE connects to the BS that provides maximum long-term average received power. In Fig.~\ref{fig:1tier1} (left), the typical UE connects to its nearest BS, which is the one that corresponds to the Voronoi cell in which it lies. Denote the location of this serving BS by $x_s \in \Phi$. The second step is to independently thin the point process $\Phi \setminus x_s$ where each point is independently retained with probability $p$ as shown in Fig.~\ref{fig:1tier1} (middle). Since the thinning is conditional on the serving BS, we call it {\em conditional thinning}. The thinned process $\Phi_b$ models the BS locations. Due to conditional thinning, the point process $\Phi_b$ is no longer a PPP. After tessellating the space based on $\Phi_b$ and keeping the position of the UE fixed, we note that the UE is pushed towards the cell interior compared to the new cell edge as shown in Fig.~\ref{fig:1tier1} (right). As discussed in the next section, a typical UE can be pushed arbitrarily close to its serving BS by choosing an arbitrarily small value of $p$ but remains edge biased for high values of $p$. 

\section{Impact on Cellular Networks}
After providing an overview of the new UE sampling model, we now formalize the effect of conditional thinning perceived by a typical UE. Denoting $R_1$ and $R_2$ to be the distances of the closest and the next closest point of $\Phi_b$ to the origin, we define our first metric to be $R = R_2/R_1$. It corresponds to the ratio of the distances of a typical UE to its serving BS and the dominant interferer, and provides some insights into the expected performance of a typical UE. For instance, if the value of $R$ is close to $1$, it means that the UE is near the cell edge, i.e., the dominant interferer is approximately at the same distance as the serving BS and hence the received signal-to-interference-ratio ($\sir$) is expected to be low. It is important to note that $R \geq 1$ by construction, since the serving BS is always the closest one in our cell association model. We now derive the distribution of $R$ after deriving the joint distribution of $R_1$ and $R_2$ as a function of thinning probability $p$ in the following Lemma. Interested readers can refer to~\cite{HaeJ2005} for the marginal distribution of $R_n$.

\begin{lemma}
The cumulative distribution function (CDF) of R is
\begin{equation}
F_R(r) = 1 - \frac{1}{1+ p(r^2-1)},\ r\geq 1.
\end{equation}
\end{lemma}

\begin{IEEEproof}
For notational simplicity, we first define the following disjoint sets:
\begin{align}
\calE_1 & = \{x \in \R^2: \|x\| \leq r_1\} \nonumber \\
\calE_2 & = \{x \in \R^2: r_1 < \|x\| \leq r_1 + \dd r_1   \} \nonumber \\
\calE_3 & = \{x \in \R^2: r_1 + \dd r_1 < \|x\| \leq r_2   \} \nonumber \\
\calE_4 & = \{x \in \R^2: r_2 < \|x\| \leq r_2 + \dd r_2   \}, \nonumber
\end{align}
where $\calE_1$ denotes a circle centered at origin and the rest denote annular regions defined by concentric circles centered at the origin. Further let $N(\calE)$ be a random counting measure of a Borel set $\calE$, i.e., $N(\calE) = \#$ of points in $\calE$. Now to derive the CDF of $R$, we first derive the joint probability density function (PDF) of $R_1$ and $R_2$, which by definition can be expressed as
\begin{align}
f_{R_1, R_2} (r_1, r_2) &= \lim\limits_{\substack{\dd r_1 \rightarrow 0 \\ \dd r_2 \rightarrow 0} } 
\frac{\P\left( R_1 \in \calE_2, R_2 \in \calE_4 \right)}{\dd r_1 \dd r_2}.
\label{eq:jpdf_def}
\end{align}
The numerator of the above expression can be expressed as:
\begin{align}
&\P\left( R_1 \in \calE_2, R_2 \in \calE_4 \right) = \nonumber \\
&\P\left( N(\calE_1) = 0, N(\calE_2) = 1, N(\calE_3) = 0, N(\calE_4)=1 \right) =\\
&\P\left( N(\calE_1) = 0 \right) \P\left( N(\calE_2) = 1 \right) \P\left( N(\calE_3) = 0 \right) \P\left( N(\calE_4)=1 \right), \nonumber
\end{align}
where the simplification follows from the fact that the sets $\calE_i$ are disjoint. Now recall that the point process $\Phi$ is independently thinned by probability $p$ outside the circle of radius $r_1 + \dd r_1$. Therefore, the above expression can be written as:
\begin{align}
&\P\left( R_1 \in \calE_2, R_2 \in \calE_4 \right) =
e^{-\lambda \pi r_1^2} \nonumber \\
&\lambda \pi \left[(r_1 + \dd r_1)^2 - r_1^2\right] e^{-\lambda \pi \left[(r_1 + \dd r_1)^2 - r_1^2\right]} e^{-p \lambda \pi \left[r_2^2 - (r_1 + \dd r_1)^2\right]}  \nonumber \\
&p \lambda \pi \left[(r_2+\dd r_2)^2 - r_2^2 \right]
e^{-p \lambda \pi \left[(r_2+\dd r_2)^2 - r_2^2\right]},
\end{align} 
which for vanishingly small $\dd r_1$ and $\dd r_2$ can be simplified to:
\begin{align}
%\P\left( R_1 \in \calE_2, R_2 \in \calE_4 \right) = 
p (2 \pi \lambda)^2 r_1 r_2 \exp\left(-\lambda \pi r_1^2 (1-p)\right) \exp\left(-p \lambda \pi r_2^2\right) \dd r_1 \dd r_2,
\end{align}
from which the joint PDF of $R_1$ and $R_2$ can be expressed as:
\begin{align}
f_{R_1, R_2} (r_1, r_2) =  p (2 \pi \lambda)^2 r_1 r_2 e^{-\lambda \pi r_1^2 (1-p)} e^{-p \lambda \pi r_2^2},
\label{eq:pdfR1R2}
\end{align}
for $r_2 \geq r_1 \geq 0$. Using this joint density, the CCDF of $R$ can now be expressed as:
\begin{align}
&\P[R > r] = \P \left[\frac{R_2}{R_1} > r \right] \\
&= \int_{r_2 = 0}^{\infty} \int_{r_1 = 0}^\frac{r_2}{r} f_{R_1, R_2} (r_1, r_2) \dd r_1 \dd r_2
%&= p (2 \pi \lambda)^2  \int_{r_2 = 0}^{\infty}  r_2 \exp\left(-p \lambda \pi r_2^2\right)  
%\int_{r_1 = 0}^{\frac{r_2}{r}} r_1 \exp\left(-\lambda \pi r_1^2 (1-p)\right)  \dd r_1 \dd r_2 \\
= \frac{1}{1 + p(r^2 - 1)},
\end{align}
which completes the proof.
\end{IEEEproof}

Recall that for $p=1$ the proposed model reduces to the uniform distribution of UEs, leading to the following corollary. 
\begin{cor}
The CDF of $R$ for the typical observation point that is defined to be uniformly distributed in $\R^2$ independent of the BS locations ($p=1$ in the proposed model) is
\begin{align}
F_R(r) = 1 - \frac{1}{r^2},\ r\geq 1.
\end{align}
\end{cor}

\begin{cor} \label{cor:Rmean}
The mean value of $R$ as a function of $p$ is
\begin{align}
\E[R] = 1+ \frac{1}{\sqrt{p(1-p)}} \left(\frac{\pi}{2} - \tan^{-1} \left(\frac{\sqrt{p}}{\sqrt{1-p}} \right) \right),
\end{align}
which for $p=1$ is $\E[R] = 2$.
\end{cor}

\begin{figure}[t]
\centering
\includegraphics[width=\columnwidth]{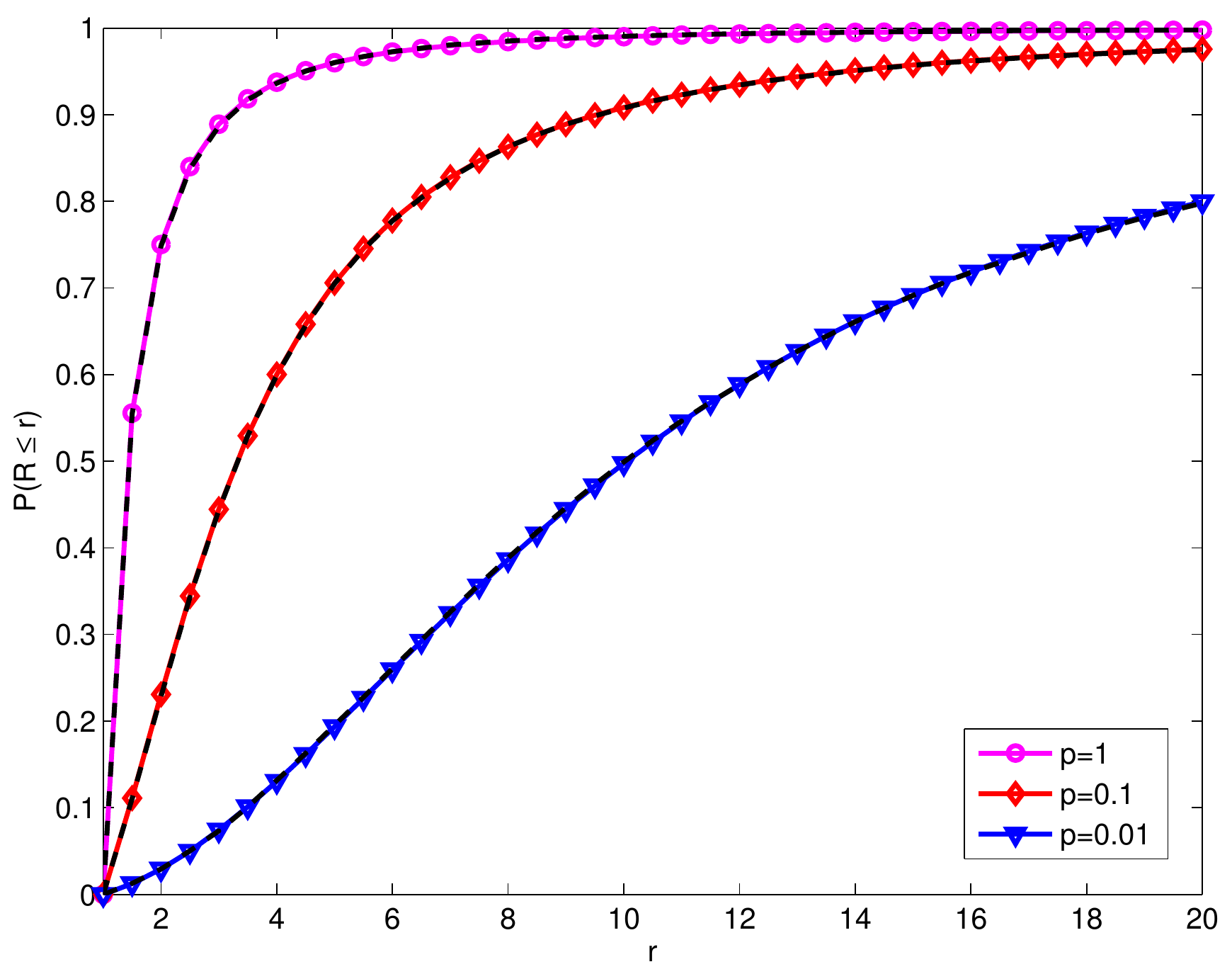}
\caption{The CDF of $R$ for various values of conditional thinning probability $p$. The theoretical results are overlaid with dotted plots of the simulation results showing perfect match.}
\label{fig:R_CDF}
\end{figure}

\begin{remark}[Dependence in UE and BS point processes]
The level of dependence induced in the locations of the UEs and the BSs is inversely proportional to the thinning probability $p$, i.e., the probability of finding a typical UE in the cell interior close to its serving BS is higher for smaller values of $p$. This is evident from the mean of $R$ given by Corollary~\ref{cor:Rmean} and from the plot of the CDFs of $R$ for various values of $p$ in Fig.~\ref{fig:R_CDF}. %Also note that in case of $p=1$, the dominant interferer is on an average only twice as far as the serving BS, suggesting that the performance estimates derived using this model are fairly edge biased. This is expected under uniform UE distribution assumption because more users lie near the cell edge than in the cell interior.
\end{remark}

We now derive the coverage probability of a typical UE sampled under the new proposed method. The coverage probability $\pc$ denotes the average fraction of UEs in coverage and can be formally defined as the CCDF of $\sir$ as follows:
\begin{align}
\pc = \P\left( \sir > \T \right)
= \P \left( \frac{h_{x_s} \|x_s\|^{-\alpha}} {\sum_{y \in \Phi_b \setminus x_s}  h_y \|y\|^{-\alpha}} > \T  \right),
\end{align}
where the serving BS is assumed to be located at $x_s$. Using tools developed in~\cite{AndBacJ2011}, a simple expression can be derived for the coverage probability of a typical UE under the proposed sampling method. The main result is given in the following Lemma along with a brief proof sketch.

\begin{lemma} \label{lem:coverage}
The coverage probability of a typical UE when it connects to its nearest BS under the proposed UE sampling method with thinning probability $p$ is
\begin{align}
\pc (\alpha, \T, p) = \left[ 1+ p  \T^\frac{2}{\alpha} \int_{\T^{-\frac{2}{\alpha}}}^{\infty} \frac{1}{1+u^\frac{\alpha}{2}} \dd u \right]^{-1},
\end{align}
which for $\alpha = 4$ simplifies to
\begin{align}
\pc (4, \T, p) = \left[1 + p \T^{\frac{1}{2}} \left(\pi/2 - \arctan (\T^{-\frac{1}{2}}) \right) \right]^{-1}.
\end{align}
%and when it connects to the strongest BS in terms of received power (assuming $\T>1$) is
%\begin{align}
%\pc(\alpha, \T, p)=& \frac{\pi \T^{-2/\alpha}}{C(\alpha)} -\sum_{m=1}^\infty g(m),
%\label{eq:Pc_sameT}
%\end{align}
%where,
%\begin{align}
%g(m)=\left(\frac{-\pi\Gamma(1+\frac{2}{\alpha})(1-p)}{C(\alpha)p \T^{2/\alpha}} \right)^m
%\left\{\frac{1}{\Gamma(1+\frac{2m}{\alpha})}-\frac{\pi\Gamma(1+\frac{2}{\alpha}) {}_2F_1(1,\frac{2m}{\alpha},1+\frac{(m+1)2}{\alpha},\frac{1}{1+\T})}{C(\alpha)\Gamma(1+\frac{(m+1)2}{\alpha})\T^{2/\alpha}(1+\beta)^{2m/\alpha}}\right\},
%\end{align}
%and $C(\alpha) =  \frac{2 \pi^2 \csc \left(\frac{2 \pi}{\alpha} \right)}{\alpha}$ and the hypergeometric function is denoted by
%\begin{align}
%{}_2F_1(a,b,c,z)=\frac{\Gamma(c)}{\Gamma(b)\Gamma(c-b)}\int_0^1\frac{t^{b-1}(1-t)^{c-b-1}}{(1-tz)^a}\dd t.
%\end{align}
\end{lemma}
%\begin{IEEEproof}

This result directly follows from Theorem 2 of~\cite{AndBacJ2011} with a slight modification to incorporate conditional thinning of the point process $\Phi$. %The required modification is explained in the following proof sketch. 
The coverage probability is first conditioned on the distance of a typical UE to its serving BS, which can be computed by the null probability of a PPP. Conditioned on this distance, say $u$, the interference field defined over $\R^2 \cap B(0,u)^c$ is a PPP with density $p\lambda$, where $B(0,u)$ is a ball with radius $u$ centered at $0$ and $p$ appears due to conditional thinning. This is the step where the proof of the current Lemma differs from that of Theorem 2 of~\cite{AndBacJ2011}. The remaining proof, which mainly involves the derivation of the Laplace transform of interference, remains the same.

\begin{remark}[Scale invariance and effect of $p$ on $\pc$]
For any given value of thinning probability $p$, the coverage probability is independent of the density of the BSs in an interference limited cellular network. This scale invariance result is a generalization of a similar observation reported in~\cite{AndBacJ2011} for the uniform distribution of the UEs, which is a special case of the proposed model and corresponds to $p=1$. As expected, the coverage probability is a monotonically decreasing function of $p$. This is consistent with the observations reported in~\cite{DhiGanJ2012a}, where a similar parametrization was used to model load on various classes of BSs in a heterogeneous cellular network.
\end{remark}

\begin{figure}[ht!]
\centering
\includegraphics[width=.66\columnwidth]{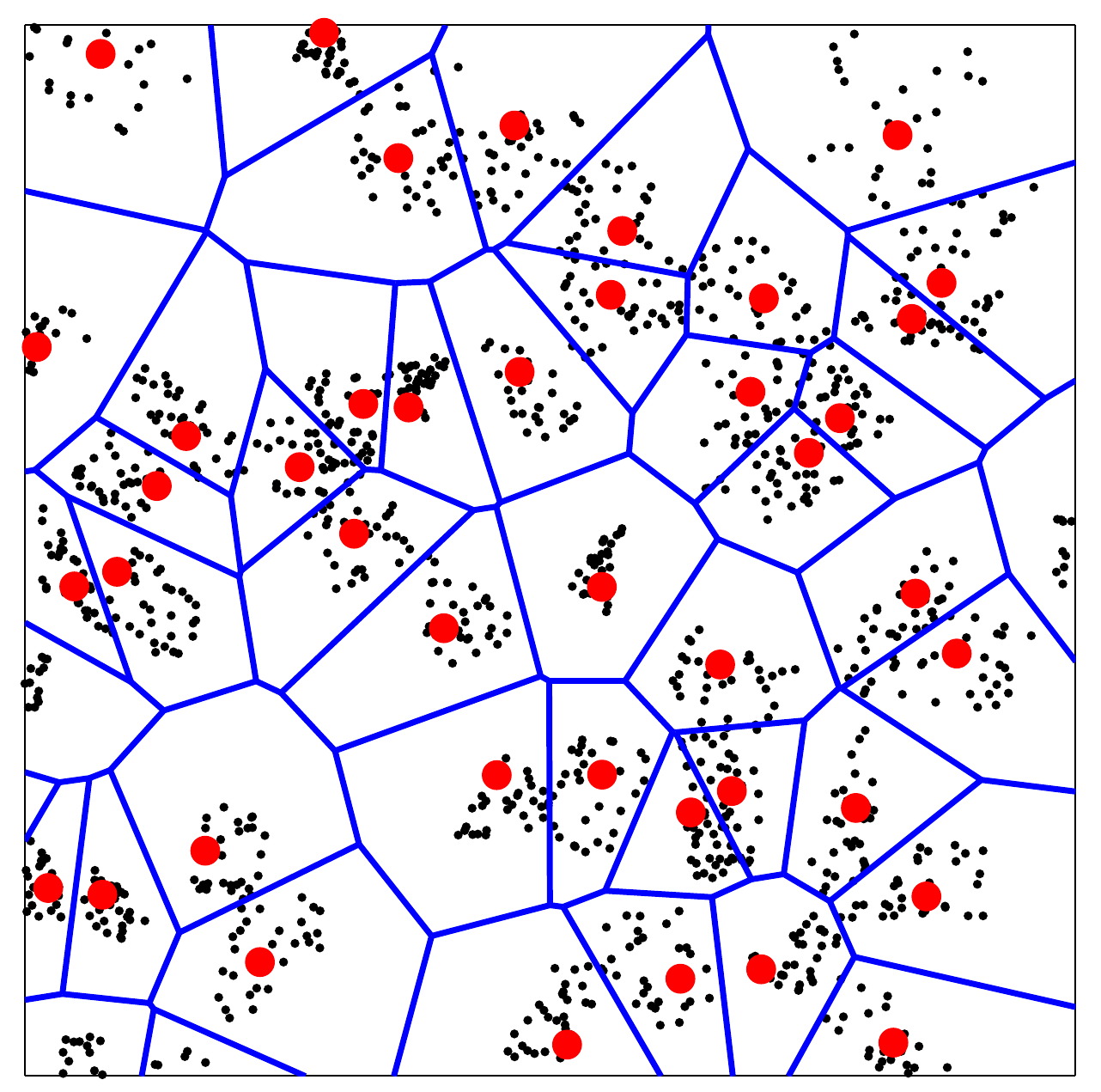}
\caption{A realization of the proposed non-uniform UE distribution model.}
\label{fig:UE_nonunif}
\end{figure}

\section{Non-Uniform UE Distribution}
In this section, we show that the framework developed in the previous section can be used as an accurate analytical generative model for the non-uniform UE distributions where the UEs are clustered around their serving BSs. We first explain the simulation setup for this non-uniform UE distribution model and then describe a subtle difference in this model and the UE sampling framework developed in the previous section.

To simulate the non-uniform UE distribution model, start with a realization of a PPP $\Phi$ with density $\lambda$ and form the Voronoi tessellation. Recall that in case of a cellular network, the Voronoi cell of each point denotes its coverage region in the nearest neighbor connectivity model. Distribute $N_u$ users uniformly in each Voronoi cell. For concreteness, we assume that $N_u$ is the same for all the BSs and equal to the number of resource blocks, which models a full-buffer system. This assumption can be relaxed under certain conditions as discussed in~\cite{DhiGanJ2012a}. Until this point, the UE distribution is fairly uniform although there is a subtle difference in this model and the way typical user is usually defined to be uniformly distributed over $\R^2$. We will remark on this difference later in this section. To induce dependence in the BS and UE point processes, we again use the the thinning idea and retain points of the realization of $\Phi$ independently with probability $p$ and remove the rest. The UEs corresponding to the points that are removed are also removed. The thinned version of the point process $\Phi$ with density $\lambda p$ models the BS locations. For the new coverage areas defined by the thinned point process, the remaining UEs are biased towards the cell interior as shown in Fig.~\ref{fig:UE_nonunif}. A favorable characteristic of this model is the probabilistic attraction introduced in the UE and BS point processes without inducing any geometric constraints. %This is evident in Fig.~\ref{fig:UE_nonunif}, where we note that while the UE locations are biased towards the BS locations over the whole network, they are highly concentrated around certain BSs and relatively more spread out for others. 
The coverage probability can now be numerically evaluated by averaging over these UE locations. Note that the result of Lemma~\ref{lem:coverage} is not exact for this simulation model, as remarked below.

\begin{remark}[UE uniformly distributed in $\R^2$ vs in randomly chosen Voronoi Cell] There is a subtle difference in performing downlink analysis at a typical UE uniformly distributed in $\R^2$ and uniformly distributed in a randomly chosen Voronoi cell. The former corresponds to the analytical model discussed in the previous section for $p=1$ and the latter corresponds to the simulation model discussed in this section. The difference is induced by the structure of Poisson Voronoi tessellation and can be understood by a simple fact that a point uniformly distributed in $\R^2$ is more likely to fall in a bigger Voronoi cell, whereas there is no such bias when we choose a point uniformly distributed in a randomly chosen Voronoi cell. %This is the reason why the coverage probability result given by Lemma~\ref{lem:coverage} is not exact for the simulation model described in this section, although it will be shown to provide a tight approximation.
\label{rem:difference}
\end{remark}

\begin{figure}[ht!]
\centering
\includegraphics[width=\columnwidth]{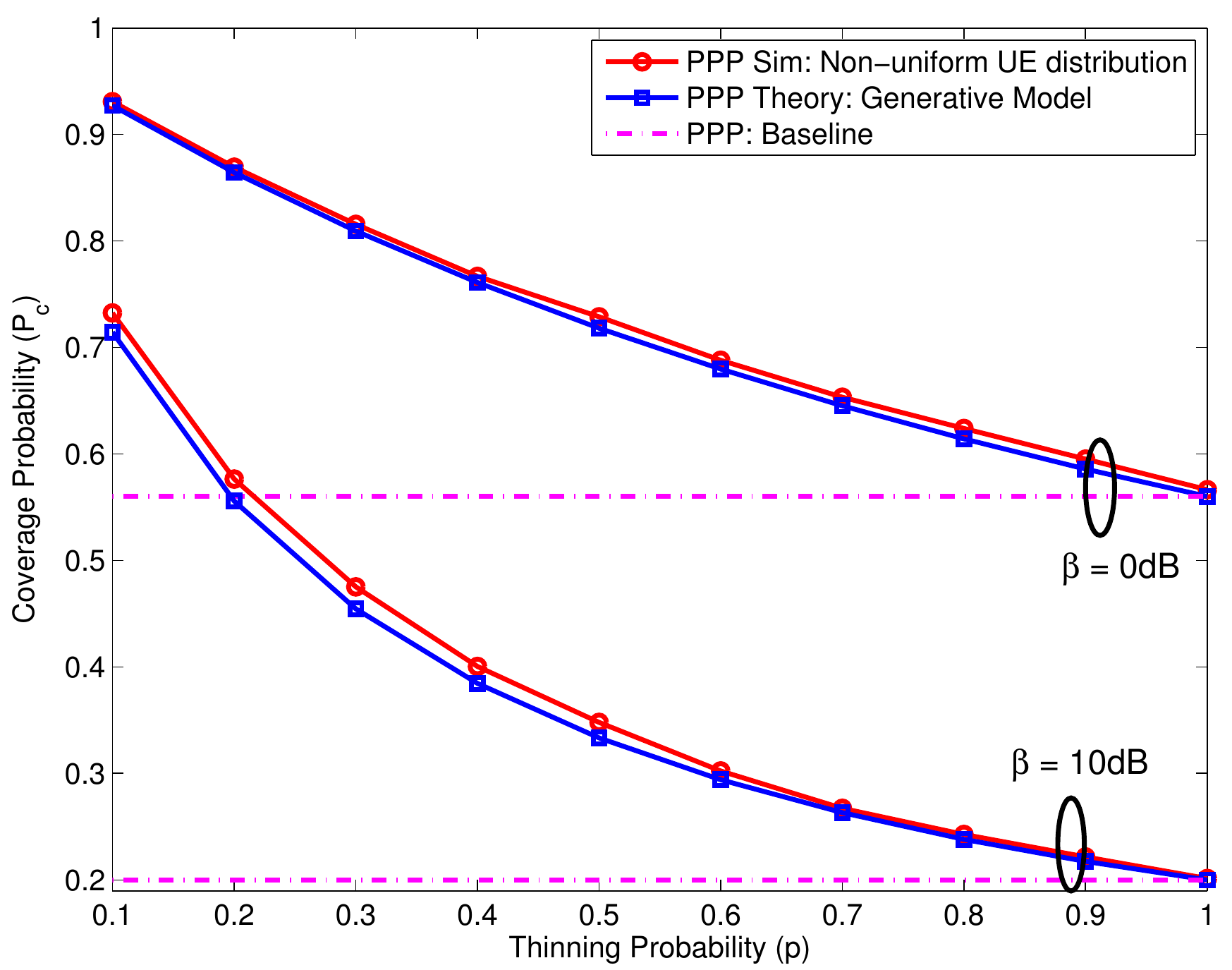}
\caption{Comparison of the coverage probability of the proposed non-uniform UE distribution model with the analytical expression derived in Lemma~\ref{lem:coverage} and the baseline model assuming uniform UE distribution ($\alpha = 4$).}
\label{fig:Pc_main}
\end{figure}

We now compare the coverage probability of this non-uniform UE distribution model with the analytical expression derived in Lemma~\ref{lem:coverage} in Fig.~\ref{fig:Pc_main}. We first note that the plots are surprisingly close and the difference highlighted in Remark~\ref{rem:difference} has a negligible impact on the coverage probability. Thus, the analytical model based on conditional thinning can be used as an accurate generative model to study coverage probability for this non-uniform UE distribution model described in this section. Further, we compare the coverage probability with the baseline model, where the UEs are distributed uniformly over $\R^2$ independent of the BS point process. We note that the difference in coverage predictions is significant even for high values of $p$. This clearly highlights the importance of accurate UE distribution models in the performance analysis of cellular networks.

\section{Conclusion}
In this letter, we addressed the problem of incorporating non-uniform UE distributions in the random spatial models for cellular networks. %Based on the idea of conditional thinning, we first proposed a model to bias the location of a typical UE towards cell interior and then used it as a generative model to study deployment scenarios where the UEs are more likely to lie close to their serving BSs.  
This work has numerous extensions, both in system modeling and analysis. In system modeling, it is important to develop empirical UE distribution models, especially for heterogeneous networks, where the UEs are more likely to be around low power nodes. In analysis, it is important to develop tractable modeling tools capable of handling correlations in the UE and BS point processes.
%In analysis, it is important to understand Palm distributions for these dependent point processes and develop tractable modeling tools to study these new deployment and usage trends to provide more realistic system design guidelines.

\bibliographystyle{IEEEtran}
\bibliography{WCL_NonUnif_UE_ArXiv_v1}

\end{document}